\documentclass[useAMS,usenatbib,usegraphicx]{mn2e}

\title[Spatial distribution of luminous X-ray binaries in spiral galaxies]
{Spatial distribution of luminous X-ray binaries in spiral
galaxies}
\author[Z. Zuo et al.]
{Zhao-yu Zuo $^{1,2}$\thanks{E-mail:zhaoyuz@astro.umass.edu},
Xiang-dong Li$^1$\thanks{lixd@nju.edu.cn} and Xi-wei
Liu$^{1,3}$\thanks{liu..xw@163.com}\\
$^1$Department of Astronomy, Nanjing University, Nanjing 210093, China\\
$^2$Department of Astronomy, University of Massachusetts, Amherst, MA 01003, USA\\
$^3$Department of Physics, Central China Normal University, Wuhan
430079, China}
\begin{document}
\date{Accepted  . Received  ; in original form }

\pagerange{\pageref{firstpage}--\pageref{lastpage}} \pubyear{2002}

\maketitle

\label{firstpage}

\begin{abstract}
We have modeled the spatial distribution of luminous X-ray
binaries (XRBs) in Milky Way like spiral galaxies with an
evolutionary population synthesis code developed by
\citet{Hurley00,Hurley02}. In agreement with previous theoretical
expectations and observations, we find that both high- and
low-mass X-ray binaries show clear concentrations towards the
galactic plane and bulge. We also compare XRB distributions under
the galactic potential with dark matter halo and the Modified
Newtonian Dynamics (MOND) potential, and suggest that the
difference may serve as potential evidence to discriminate these
two types of models.

\end{abstract}


\section{Introduction}

X-ray binaries (XRBs) contain a neutron star (NS) or a black hole
(BH) accreting from a normal companion star. They are conventionally
divided into low-mass X-ray binaries (LMXBs) and high-mass X-ray
binaries (HMXBs) according to the masses of the optical companions
\citep[e.g.][]{verbunt94}. In HMXBs, the evolved (super)giant
companions, generally $M_{\rm optical}\ga 10M_{\odot}$, have strong
stellar wind mass-loss to power a  bright X-ray source for $\sim
10^{5}-10^{6}$ yr; while LMXBs, in which $M_{\rm optical}\la 1.5
M_{\odot}$, experience mass transfer via Roche-lobe overflow (RLOF)
at a rate of $\sim 10^{-10}-10^{-8} M_{\odot}$ yr$^{-1}$. Between
them are intermediate-mass X-ray binaries (IMXBs), in which
companion stars' masses are in the range $\sim 2 - 10 M_{\odot}$
\citep{heuvel75}. Mass transfer in these binaries often occurs on a
(sub)thermal timescale of $\sim 10^4-10^5$ yr through RLOF.

Using distance estimates and angular distribution of LMXBs,
\citet{paradijs95} and \citet{white96} investigated the spatial
distribution of NS and BH LMXBs in our Galaxy, and suggested that
the compact objects had received a kick during the supernova (SN)
explosions. More recent work by \citet{grimm02} using the {\it
RXTE\/} data showed that HMXBs were concentrated towards the
Galactic plane with a vertical scale height of $\sim 150$ pc while
the vertical distributions of LMXBs was significantly broader with
a scale height of $\sim 410$ pc, and the radial distribution of
LMXBs peaked strongly at the Galactic bulge. But this sample
suffers from some incompleteness of the optical
identifications/distance measurements at the large distances from
the Sun \citep{jonker04}. Fortunately, today's sensitive,
high-resolution X-ray observations allow the study of luminous
XRBs in galaxies even beyond the Local Group, and make it possible
to examine XRB populations in a wide range of galactic
environments with different star formation histories. For example,
{\it XMM-Newton\/} and {\it Chandra\/} observations of NGC 891, a
nearby edge-on spiral galaxy which is very similar in many
respects to our own Galaxy, present a straightforward look from
outside \citep{temple05}. The spatial distribution of luminous
discrete point sources in this galaxy also shows clear
concentrations towards the galactic plane and bulge. From the
locations of 154 discrete non-nuclear ultraluminous X-ray sources
(ULXs) identified in 82 galaxies observed with {\it Chandra\/},
\citet{swartz04} found that the ULXs in their host galaxies were
strongly peaked toward their galaxy centers. Statistical analysis
of the X-ray point sources from the {\it ROSAT HRI} survey of
nearby galaxies by \citet{liujf06} showed that there is a
significant concentration of ULXs towards galactic center in
late-type galaxies. They also suggested that regular ULXs are
likely to be a high-luminosity extension of the ordinary HMXB/LMXB
population in late-type galaxies through luminosity function (LF)
study.

The spatial distribution of XRBs in a galaxy is determined by the
initial kick velocity due to any asymmetry in SN explosion at the
birth of a NS/BH, the galactic gravitational potential, and the
mass transfer process in a binary. In the present work, we
investigated the dynamical consequences of XRBs in spiral galaxies
like the Milky Way in an theoretical view. We employed an
evolutionary population synthesis (EPS) code to calculate the
expected number and luminosity distributions of XRBs in the
galaxies. Then, following the approach of \citet{paczynski90}, we
calculated the spatial distribution of XRBs with luminosities
$>10^{37}$ ergs$^{-1}$. For the galactic gravitational potential,
we adopted both the standard cold dark matter (CDM) model and the
Modified Newtonian Dynamics (MOND) model. The objective of this
study is to present an integrated picture of XRB distribution in
spiral galaxies under the two kinds of galactic potential models,
and to explore the difference in the predicted spatial
distributions, which could be testified by comparison with future
high-resolution observations of XRB distribution in nearby
galaxies. A recent related work is to use the detection of LMXBs
in the Sculptor dwarf spheroidal galaxy to probe the dark matter
halo \citep{dehnen06}.

This paper is organized as follows. In \S 2 we describe the
population synthesis method and the input physics for XRBs in our
model. The calculated results are presented in \S 3. Our discussion
and conclusions are in \S 4.

\section{Model}

\subsection{Assumptions and input parameters}

We have used the EPS code developed by \citet{Hurley00,Hurley02} to
calculate the expected numbers for various types of XRB populations.
This code incorporates evolution of single stars with binary-star
interactions, such as mass transfer, mass accretion, common envelope
(CE) evolution, SN kicks, tidal friction and angular momentum loss
mechanics (i.e., mass loss, magnetic braking and gravitational
radiation). Besides the modifications made by \citet{liu07} to the
original code, we have reduced the helium star wind strength by a
factor of 0.6 according to \citet{Kiel06} in modeling the formation
processes of BH LMXBs.

We assume that the host spiral galaxies are similar to our Galaxy,
and adopt the cylindrical coordinate system ($R$, $\phi$, and $z$)
centered at the galactic center. For stars born in the bulge, we
simply assume that they were distributed uniformly between $R_{\rm
min}=0$ and $R_{\rm max}=2$ kpc, while the star formation rate
(SFR) in the disk varies exponentially with $R$, i.e., $\propto
\exp(-R/R_{\rm exp})$ with $R_{\rm exp}=4.5$ kpc, from $R_{\rm
min}=2$ out to $R_{\rm max}=15$ kpc.

Considering the different star formation processes in the galactic
disk and bulge (described in \S 3.1), we have calculated the
populations of X-ray sources in the bulge and disk separately.
According to \citet{ballero07} we assume that stars in the bulge
formed at the age of 0.4 Gyr with metallicity of 0.001 and an
inital mass function (IMF) more skewed toward high mass than in
the solar neighbourhood. We neglect magnetic braking effect for
main-sequence (MS) stars of mass $0.8-1.25 M_{\odot}$, since
metal-poor MS stars in this mass range do not have an outer
convective zone \citep[e.g.][]{ivanova06a}. For the disk we take a
fixed SFR over the lifetime of the galaxy and solar metallicity.
The values of other adopted parameters are the same as the default
ones in \citet{Hurley02} if not mentioned. The IMF of
\citet{Kroupa} is taken for the primary's mass ($M_1$)
distribution. For the secondary stars ($M_2$), we assume a uniform
distribution of the mass ratio $M_2/M_1$ between 0 and 1 and of
the logarithm of the orbital separation $\ln a$. Tidal effect is
considered to remove any eccentricity induced in a post-SN binary
prior to the onset of mass-transfer.

When a binary survives a SN explosion, it receives a velocity kick
due to any asymmetry in the explosion \citep{lyne94}. The kick
velocity $v_{\rm k}$ is assumed to be imparted on the newborn NS
with the Maxwellian distribution
\begin{equation}
   P(v_{\rm k})=\sqrt{\frac{2}{\pi}}\frac{v^{2}_{\rm k}}{\sigma^{3}}
   exp(-\frac{v^{2}_{\rm k}}{2\sigma^{2}})
\end{equation}
where $\sigma=265{\rm kms}^{-1}$ \citep{Hobbs} or 190
kms$^{-1}$\citep{hansen97}. The direction of the initial velocity
vector is chosen randomly. Together with the local circular motion
in our Galaxy \citep{burton78}, this gives the initial velocity
vectors $v_{\rm R}$, $v_{\phi}$, $v_{\rm z}$. After evolving for a
period, the binary will turn on X-rays and can be observed if it
is luminous enough.

In the mean time the motion of the binary can be calculated if the
galactic potential is known. In our control model we adopt the
Galactic gravitational potential proposed by \citet{johnston95},
which consists of one Hernquist bulge \citep{hernquist90}, one
Miyamota-Nagai disk \citep{miyamoto75} and one isothermal DM halo
potential\footnote{We have adopted other galactic potential
suggested in \citet{paczynski90} and \citet{binney87}, and found
very small difference in the final results.}. For the MOND
potential we use the \citet{shan08} model which applies MOND
correction to a Kuzmin-Hernquist bulge-disk model to study orbits
in axisymmetric potential. The potential can be constructed as,
\begin{equation}
   \Phi_{N}(R,z)=\frac{-GM}{\sqrt{R^2+(a+|z|)^2}+h}
\end{equation}
where $G$ is the gravitational constant, $a$ the Kuzmin length,
$h$ the Hernquist length and $M$ the total mass of the lens
system. Here we adopt $M=1.2\times10^{11} M_{\odot}$, $a=4.5$ kpc
\citep{binney87}, and $h=0.7$ kpc \citep{hernquist90} in our
calculations. Note that when $h\rightarrow0$, Eq.~(2) recovers to
the thin Kuzmin disk model with Newtonian potential given by
\citet{binney87}, and when $a=0$, it becomes the Hernquist model
\citep{hernquist90}. The modified gravity is then
\begin{equation}
    g=g_{N}+\sqrt{g_{N}a_{0}}
\end{equation}
where $g_N$ is calculated from Eq.~(2) and $a_0=1.2\times10^{-8}$
cms$^{-2}$ \citep{milgrom83,mcgaugh04}. For comparison we also
performed calculations with the Kuzmin model presented by
\citet{read05}.

Due to cylindrical symmetry of the galactic potential, two space
coordinates $R$ and $z$ are sufficient to describe the XRB
distributions. We integrate the motion equations \citep[i.e.,
Eqs.~(19a, b) in][]{paczynski90} with a fourth-order Runge-Kutta
method to calculate the trajectories of the binary systems and
collect the space parameters of current XRBs. In our calculations,
the accuracy of integral is set to be $10^{-6}$ and controlled by
the energy integral.

\subsection{X-ray luminosity and source type}
In our study XRBs are simply divided into LMXBs and H/IMXBs
according to the mass of the optical companion, and we use the mass
of the secondary, $M_2$, of $2 M_{\odot}$ to separate them.
Typically the donor star is a MS star but giant and white dwarf (WD)
donors are also possible. For every accreting system, the bolometric
luminosity ($L_{\rm bol}$) is calculated based on the average mass
accretion rate ($\dot{M}_{\rm acc}$) as $L_{\rm
bol}=\eta\dot{M}_{\rm acc}c^2$ where $\eta$ is the efficiency for
energy conversion and $c$ is the velocity of light. For persistent
XRBs, we adopt $L_{\rm bol}=\min(L_{\rm bol},\eta_{\rm Edd}L_{\rm
Edd})$, where $\eta_{\rm Edd}$ is the ``Begelman factor"
\citep{rappaport04} to allow super-Eddington luminosities, and the
critical Eddington luminosity $L_{\rm Edd} \simeq 4\pi GM_{1}m_{\rm
p}c/\sigma_{T}=1.3 \times 10^{38}m_{1}$\,ergs$^{-1}$ (where
$\sigma_{T}$ is the Thomson cross section, $m_{\rm p}$ the proton
mass, and $m_{1}$ the accretor mass in the units of solar mass). We
assume $\eta_{\rm Edd}=1$ and 10 for accreting NSs and BHs,
respectively. For transient sources, the luminosities in outbursts
are taken to be a fraction ($\eta_{\rm out}$) of the Eddington
luminosity. For NS systems, we assume $\eta_{\rm out}=0.1$ and 1 if
the orbital period $P_{\rm orb}$ is less and longer than 1 day,
respectively; for BH systems, we adopt $\eta_{\rm out}=P_{\rm
orb}/24$ hr and let the maximum peak luminosity not exceed $5L_{\rm
Edd}$ \citep{bel03,chen97,Garcia03}. The X-ray duty cycle (DC) is
taken to be 0.01 empirically \citep[e.g.][]{taam00}.

Finally a bolometric correction factor $\eta_{\rm bol}$ is
introduced to convert the bolometric luminosity to the X-ray
luminosity in the $2-10$ keV energy range \citep{bel03}. Generally
the correction factor is $\sim  0.1-0.5$ for different types of
XRBs, and here we adopt $\eta_{\rm bol}=0.3$ as our standard value.
So we can obtain the simulated X-ray luminosity form as follows:
\begin{eqnarray}
L_{\rm X, 2-10 kev}&=&\left\{
\begin{array} { ll}
  \eta_{\rm bol}\eta_{\rm out}L_{\rm Edd}&\ \rm transients\ in\ outbursts \\
  \eta_{\rm bol}\min(L_{\rm bol},\eta_{\rm Edd}L_{\rm Edd})&\ \rm persistent\
  systems.
\end{array}
\right.
\end{eqnarray}
To discriminate transient (t) and persistent (p) sources, we adopt
the criteria of \citet{paradijs96}  for main sequence and red
giant donors, and of \citet{ivanova06} for WD donors,
respectively.

\section{Results}
\subsection{X-ray luminosity functions (XLFs)}

We first consider the XRBs in our Galaxy. For disk sources, from
the SFR $\sim 0.25M_{\odot}$yr$^{-1}$ for stars massive than
$5M_{\odot}$ in \citet{grimm03}, we derive the total SFR in the
Galaxy following \citet{liu07}, and get the specific SFR of the
binary star population $S_{\rm b}=1.028$ yr$^{-1}$ for a binary
fraction $f=0.5$. For bulge sources, we construct a
phenomenological function of SFR(t) accordiing to Fig.~2 in
\citet{ballero07}, and assume the binary fraction $f=0.05$
\citep{ivanova05}. In Fig.~1, we show the simulated cumulative
XLFs of H/IMXBs (left) and LMXBs (middle) in the Galaxy with
$\alpha_{\rm
CE}\lambda=0.15$ \citep{dewi00}. 
Note that the XLF of H/IMXBs is significantly flatter than that of
LMXBs, as indicated in the observed XLFs derived by
\citet{grimm02}. The breaks between $\sim 10^{37}-10^{38}$
ergs$^{-1}$ in both XLFs are related to the maximum Eddington
luminosities of various types  of accreting sources (persistent NS
H/IMXBs and transient LMXBs) calculated with Eq.~(4).

\citet{grimm02} also combined the XLF of star forming galaxies in
their sample - M82, Antennae, NGC 4579, NGC 4736 and Circinus,
with a completeness limit lower than $2\times 10^{38}$
ergs$^{-1}$. These galaxies have a total SFR of $\sim 16
M_{\odot}$yr$^{-1}$, which exceeds the Galaxy SFR ($\sim 0.25
M_{\odot}$yr$^{-1}$) by a factor of $\sim 65$. They found that the
XLFs of Galactic and SMC HMXBs agree well with an extrapolation of
the combined LF of the star-burst galaxies.
We factitiously reset a SFR of $16.25 M_{\odot}$yr$^{-1}$ in our
population calculations in order to examine the effect of SFR. The
SFH is assumed to be the same as in the Galaxy ($\sim 12$ Gyr).
We show the cumulative XLFs in late-type spiral galaxies of the
right panel of Fig.~1. Note that H/IMXBs (dashed line) and LMXBs
(dotted line) dominate at the relatively high
($>10^{39}$ergs$^{-1}$) and low ($<10^{39}$ergs$^{-1}$) luminosity
in the XLF, respectively.

\subsection{Spatial distribution of XRBs in the CDM potential}

With the XLF constructed, we now calculate the spatial
distribution of H/IMXBs and LMXBs at the age of 12 Gyr. We need to
point out that we only consider the dynamical consequence of field
binaries. XRBs formed in globular clusters have a different
dynamical origin and are not included in this study. The results
are described as follows.

In Fig.~2, we show the radial distribution of luminous H/IMXBs
(left) and LMXBs (middle) in a Milky Way like galaxy. As the
figure shows, both H/IMXBs and LMXBs have a strong concentration
in the direction to the galactic center along the galactic plane
while there is a void of H/IMXBs in the galactic bulge. It is in
rough agreement with the observed distribution in our Galaxy,
considering that we have ignored the signatures of the Galactic
spiral structure. Note that H/IMXBs are dominated by disk sources
because sources in the bulge have all died early, while for LMXBs,
both bulge (dashed line) and disk (dotted line) sources contribute
to the population.

In Fig.~2 we also show the radial distributions of luminous XRBs
in late-type spiral galaxies with an enhanced SFR of $16.25
M_{\odot}$yr$^{-1}$. We only include sources with luminosities
$>10^{38}$ ergs$^{-1}$. As the figure shows, the radial
distributions also strongly cluster towards the galactic center
region and declines along the galactic plane, while there is a
marked scarcity in the galactic bulge. These features can be
compared with \citet{liujf06}, who find that for the late-type
galaxies, the radial distribution of detected ULXs shows a peak
around $0.5R_{25}$ (with $R_{25}$ as the elliptical radius of the
$D_{25}$ isophote.), and the surface number density of ULXs
decreases with radii until it flattens outside the $D_{25}$
isophotes.

In Fig.~3 is plotted the vertical distributions of H/IMXBs (dotted
line) and LMXBs (solid line) in the luminosity interval
$10^{37}-10^{38}$ ergs$^{-1}$ (left panel) and  $10^{38}-10^{39}$
ergs$^{-1}$ (right panel), respectively. Obviously H/IMXBs are more
concentrated towards the galactic plane than LMXBs. This is in
general accordance with \citet{grimm02}, who suggested that the tail
of high-$z$ LMXBs in the Galaxy cannot be solely due to the globular
cluster component because only three out of nine sources at $|z|>2$
kpc are located in globular clusters.

In the left panel of Fig.~4 we show the vertical distributions of
NS and BH LMXBs. They seem to be very similar, which is in general
agreement with those obtained by \citet{jonker04}. In order to
make a more accurate comparison, we define $R_{\rm X}(|z|)=N_{\rm
X}(z_{0}<|z|\leq10)/N_{\rm X}(0<|z|\leq10)$ to represent the ratio
of the number with $z_{0}<|z|\leq10$ kpc to the total number
between $|z|=0$ and $|z|=10$ kpc for a certain type (X) of XRB. In
the right panel of Fig.~4, we show the ratio $R_{\rm X}(|z|)$ of
H/IMXBs against $z_0$. The solid and dotted lines correspond to
sources in the luminosity interval $10^{37}-10^{38}$ ergs$^{-1}$
and $10^{38}-10^{39}$ ergs$^{-1}$, respectively. As shown in the
figure, the more luminous the XRBs are, the more concentrated they
are towards the Galactic plane, as expected. This can be easily
understood as follows. For a companion of mass $M$, the X-ray
lifetime $T \propto M/L_{\rm X}$. The higher $L_{\rm X}$, the
shorter $T$, and the smaller displacement from the original
position.

In Fig.~5, we show the schematic side view of XRB distribution in
the galaxy. The color-bar represents the normalized number ratio of
XRBs in the $R-z$ plane. Unfortunately, due to the uncertainties in
determining the distance to XRBs in the Galaxy
\citep[e.g.][]{jonker04}, it is difficult to numerically compare the
theoretical expectations with observations of bright XRBs in our
Galaxy. Observationally, \citet{temple05} presents a outside view in
X-rays of the nearby edge-on spiral galaxy NGC 891. The spatial
distribution of luminous discrete point sources in this galaxy also
shows clear concentrations towards the galactic plane and galactic
bulge.

We have performed calculations with varied key input parameters to
investigate their effect on the spatial distributions. The number
of HMXBs does not depend on the SFH in late-type galaxies since it
is much longer than the total duration for the formation and
evolution of HMXBs ($\la 10^7$ yr). According to \citet{liu07},
for young populations, binary formation rate and the factor of
super-Eddington accretion rate allowed can affect the XLFs most
prominently. Obviously H/IMXBs always follow the initial spatial
distributions of their progenitor stars (i.e. close to the disk)
due to their relatively short lifetimes. Another important
parameter in the evolution of close binaries is the CE efficiency
$\alpha_{\rm CE}$. Our calculations reveal that change of the
value of $\alpha_{\rm CE}$ does not significantly affect the
number of H/IMXBs but has a strong influence on LMXB population.
For small values of $\alpha_{\rm CE}$ the orbital motion of a
low-mass companion during the spiral-in process may be unable to
driven off the envelope of the massive BH/NS progenitor, resulting
in coalescence rather a compact binary. Variations in the CE
efficiency can change the relative numbers of various types of
X-ray binaries, but the main feature of  XRB distribution is
determined by the dynamical processes during the formation of a
BH/NS and the galactic potential adopted. Figure 6 shows the
radial (left) and vertical (right) distributions of LMXBs in the
Milky Way with different kick velocity dispersions. It is obvious
that there is no strong dependence of the overall spatial
distribution on the kick velocity dispersion, though it can affect
the outcome of binary evolution and the relative numbers of
various types of XRBs.

\subsection{Spatial distribution of XRBs in MOND potential}

We made similar calculations of XRB trajectories within the MOND
potential. In Fig.~7, we compare the radial distributions of XRBs
in a Milky Way like galaxy in both MOND and CDM potentials. The
solid, dotted, and thick short dashed lines represent those in
Shan-Zhao model, Kuzmin model and the CDM model, respectively.
Note that LMXBs in MOND potentials also show clear concentration
towards the Galactic bulge as in CDM model, but they are peaked at
$\sim 1-4$ kpc from the galactic center. This feature can be seen
clearly in the schematic side view (Fig.~8) of the distributions
of XRBs in the galaxy. In Shan-Zhao model (left panel) the
distribution shows a peak $\sim 2-3$ kpc from the galactic center,
while in Kuzmin model (right panel) it peaks at $\sim 3-4$ kpc
from the galactic center. The remarkable scarcity of XRBs within
the galactic bulge may provide interesting clues to falsity one of
these models through comparing XRB distributions with predictions
from the CDM and the MOND potentials.

We note the difference between these models is mainly due to the
potential discrepancy, especially the contribution from the bulge
component. The CDM model contains three components, one Hernquist
bulge, one Miyamota-Nagai disk and one isothermal DM halo potential.
It is the bulge potential that leads to the peaked distribution of
XRBs in the galactic center. The Kuzmin model has only one disk
component, which results in few XRBs within the galactic bulge. In
Shan-Zhao model, the auxilary point potential in Kuzmin model is
replaced with an auxilary Hernquist potential, that's to say, this
model contains not only Kuzmin disk potential but also Hernquist
bulge component. This bulge potential causes the distribution peak
to move inwards the bulge. We also find that in CDM model there is a
more extended radial distribution for XRBs than in the MOND model.
This is mainly caused by the potential of the DM halo, which is
thought to be reside beyond $10$ kpc from the galactic center to
account for the missing mass.

\section{Concluding remarks}

This study shows that, with current understanding of binary
evolutions and galactic structure, it is possible to investigate
both the luminosity function and spatial distribution of luminous
XRBs in nearby galaxies, although the results are subject to many
uncertainties and simplified treatments. For example, in our
calculations, only primordial binaries were considered while in
dense environments like the galactic bulge, dynamical formation
channels such as tidal capture, exchange encounters, and direct
collisions, may play an important role in binary formation and
change the distribution of the XRBs \citep{voss07}. Additionally,
we adopted a simplified radial distribution of newborn binaries in
the disk and  in the bulge. The actual initial distribution is
related to the structure of the spiral arms, the distribution and
evolution of the giant HII regions and warm CO clouds, which are
still poorly known. Finally, recent dynamical encounter of
galaxies may also lead to the prevalence of ULXs population and
make a great influence in their spatial distributions
\citep[e.g.,][]{fabbiano01,wolter03,belczynski04,colbert05,fabbiano06}.
Although detail comparison between observations and theoretical
predications is not available at present, rough agreement can be
obtained. In particular, our calculations show that XRBs in CDM
and MOND potentials may have distinct radial distribution around
the galactic bulge, suggesting a new way to constrain the nature
of DM and test the law of gravity. Our work motivates further
efforts to explore the origin of the spatial distributions of
luminous XRBs around the galactic center regions.

\section*{Acknowledgments}

We would like to thank Hong-sheng Zhao and Huan-yuan Shan for
valuable discussions on the MOND potential. We also thank Hai-lang
Dai, Wen-cong Chen, Jian-xia Cheng for useful comments and
discussions. We are also very grateful to an anonymous referee
whose comments and suggestions largely improved the clarity of
this paper. This work was supported by the Natural Science
Foundation of China under grant numbers 10573010 and 10221001.

\newpage

\begin{figure}
  \centering
  \includegraphics[width=5.5cm,height=5.5cm]{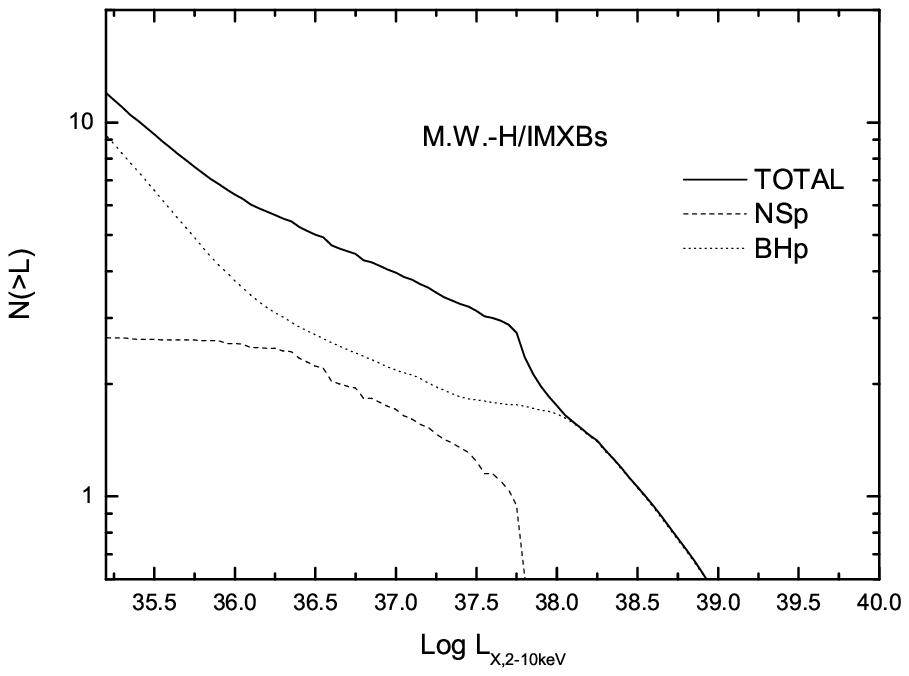}
  \includegraphics[width=5.5cm,height=5.5cm]{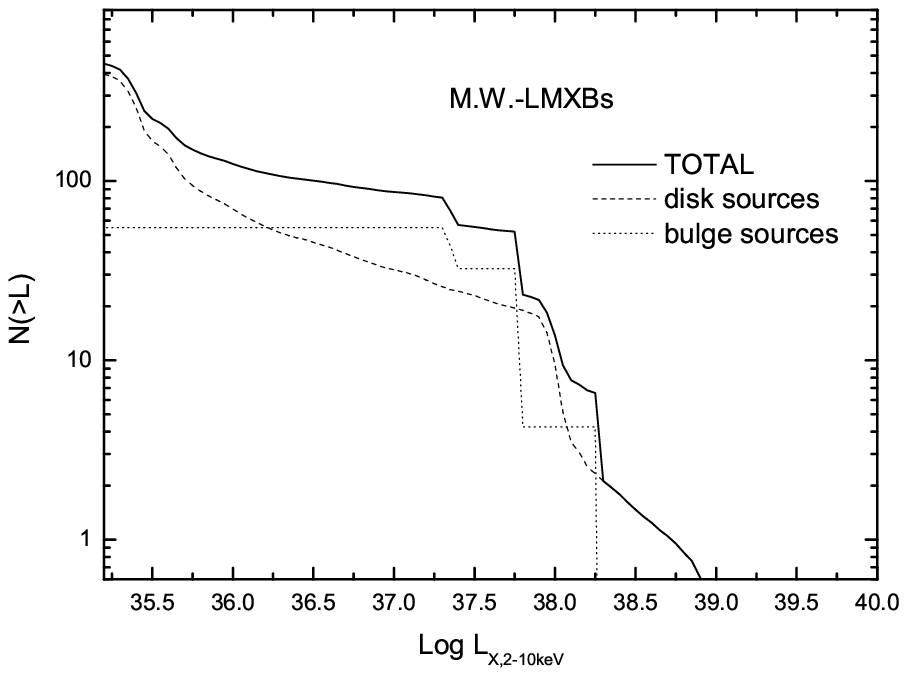}
  \includegraphics[width=5.5cm,height=5.5cm]{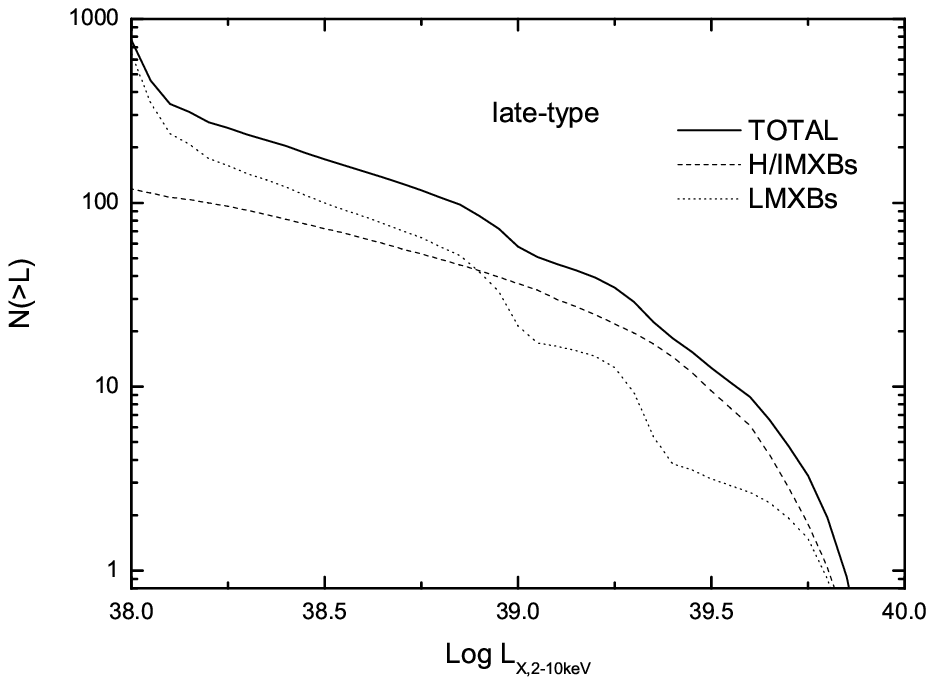}
  \caption{ Cumulative luminosity function of H/IMXBs ({\it left}), LMXBs ({\it middle}) in the Milky
  Way and luminous XRBs in late-type spiral galaxies ({\it right}). \textbf{The SFR of late-type
  spiral galaxies is 65 times of that of the Milky Way.} Note that H/IMXBs ({\it left}) dominate
  by persistent sources, both neutron star (dashed line) and black hole (dotted line) persistent sources.
  In the {\it middle} panel the dashed and dotted line represents the disk and bulge sources,
  respectively. In the {\it right} panel the thick solid line is the combined luminosity function of
  both LMXBs (dotted line) and H/IMXBs (dashed line). Note that H/IMXBs dominate at the relatively high
  ($>10^{39}$ergs$^{-1}$) luminosity while LMXBs dominate at relatively low ($<10^{39}$ergs$^{-1}$) luminosity in the
  XLF.}
  \label{Fig. 1}
\end{figure}

\newpage

\begin{figure}
  \centering
  \includegraphics[width=5.5cm,height=5.5cm]{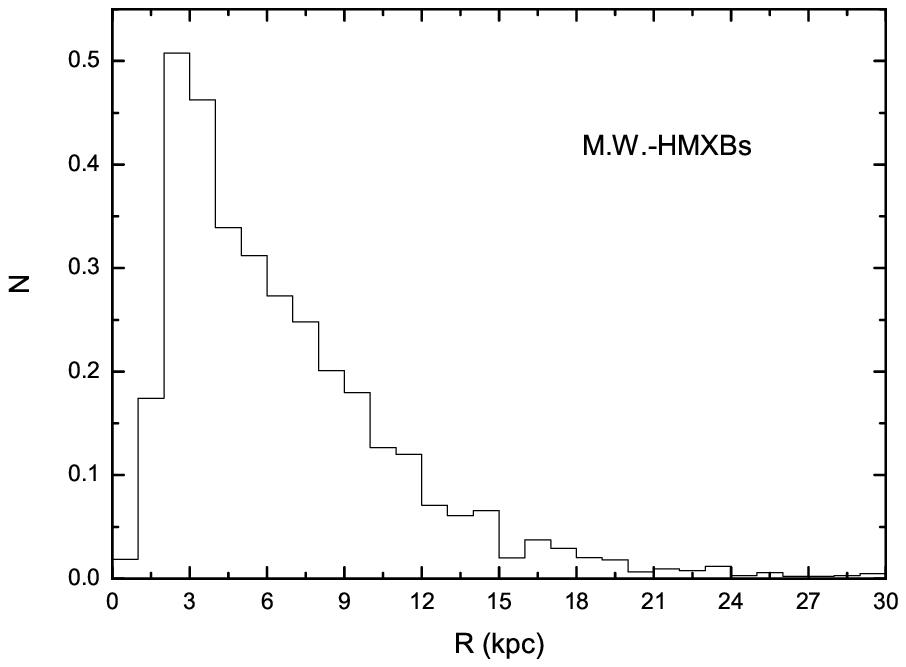}
\quad
  \includegraphics[width=5.5cm,height=5.5cm]{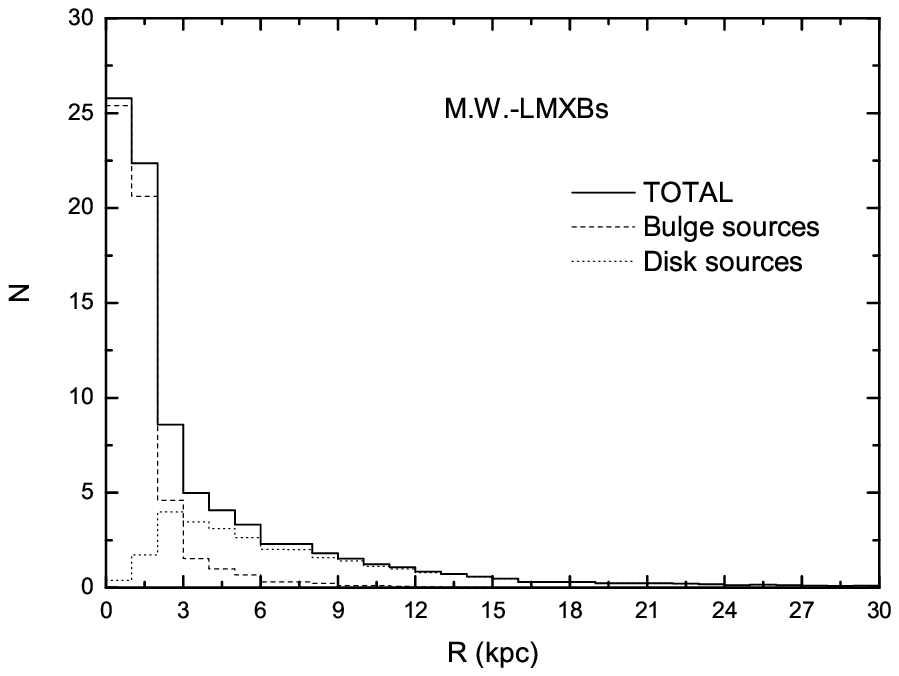}
\quad
  \includegraphics[width=5.5cm,height=5.5cm]{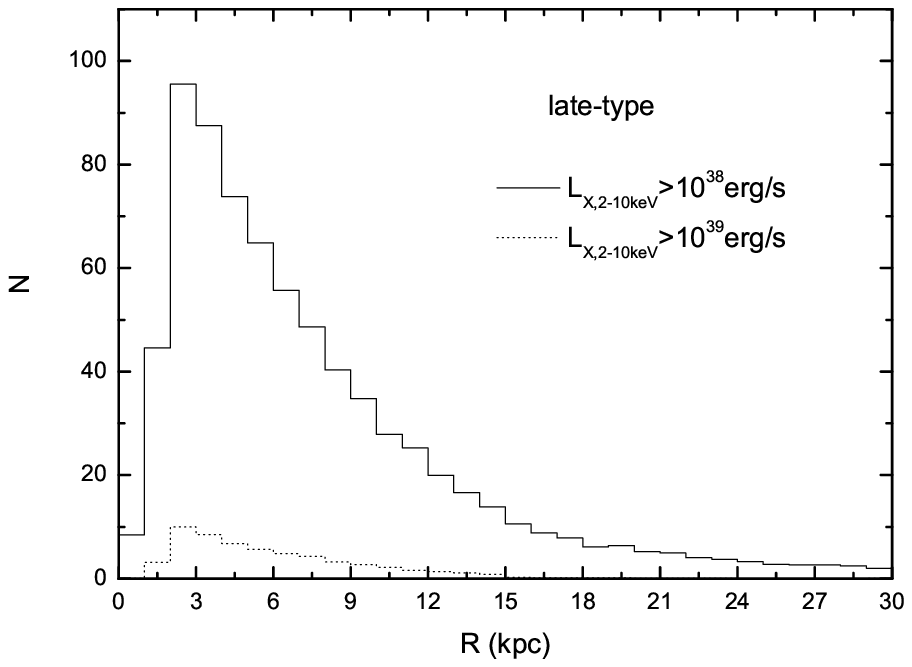}
  \caption{ Radial distributions of H/IMXBs ({\it left}), LMXBs ({\it middle}) in the Milky
  Way. Only sources with luminosity $L_{\rm x}>10^{37}$ergs$^{-1}$ are plotted.
  In the {\it right} panel shows the radial distribution of late-type spiral galaxies ( $L_{\rm
  x}>10^{38},\,10^{39}$ergs$^{-1}$). The origin of the coordinate is at the Galactic Center,
  and $R$ is the distance from the Galactic center.}
  \label{Fig. 2}
\end{figure}

\begin{figure}
  \centering
 \includegraphics[width=8cm,height=8cm]{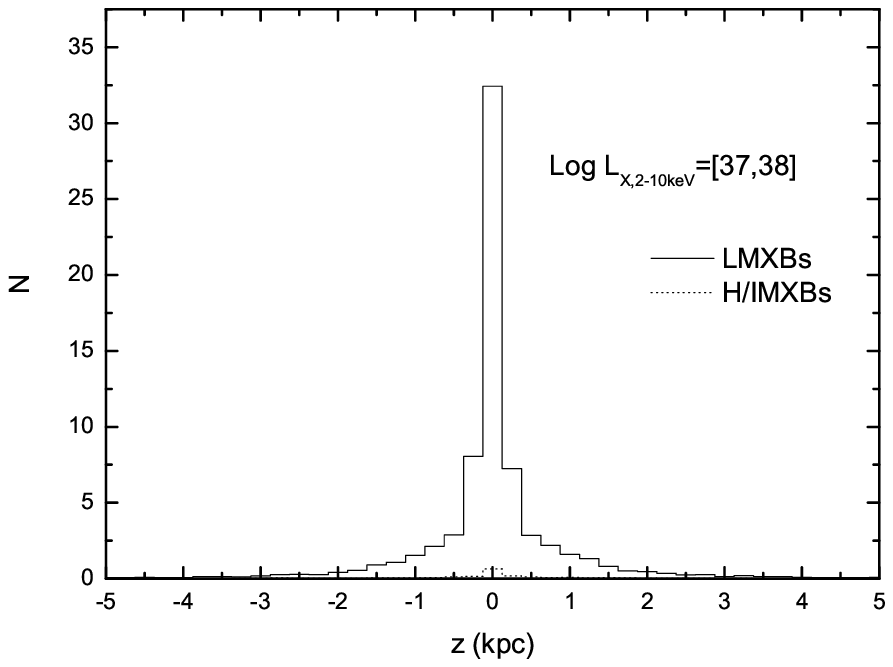}
\quad
\includegraphics[width=8cm,height=8cm]{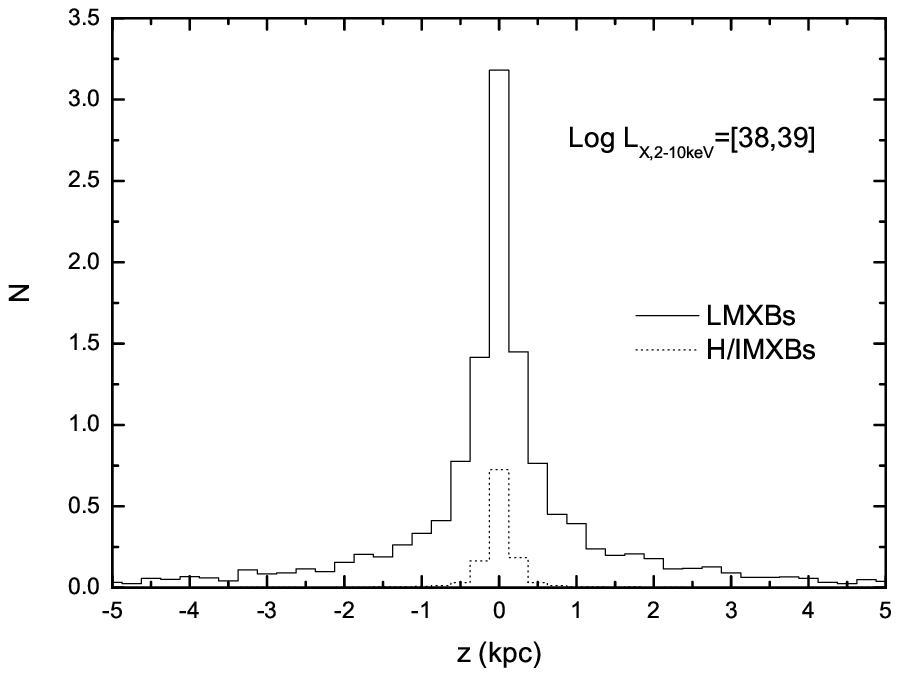}
  \caption{Vertical distributions of LMXBs (solid line) and H/IMXBs (dotted line) in the luminosity
  intervals $\log(L_{\rm X,2-10kev})=[37,38]$ (left) and $[38,39]$ (right) in the Milky Way. }
  \label{Fig. 3}
\end{figure}

\begin{figure}
  \centering
  \includegraphics[width=8cm,height=8cm]{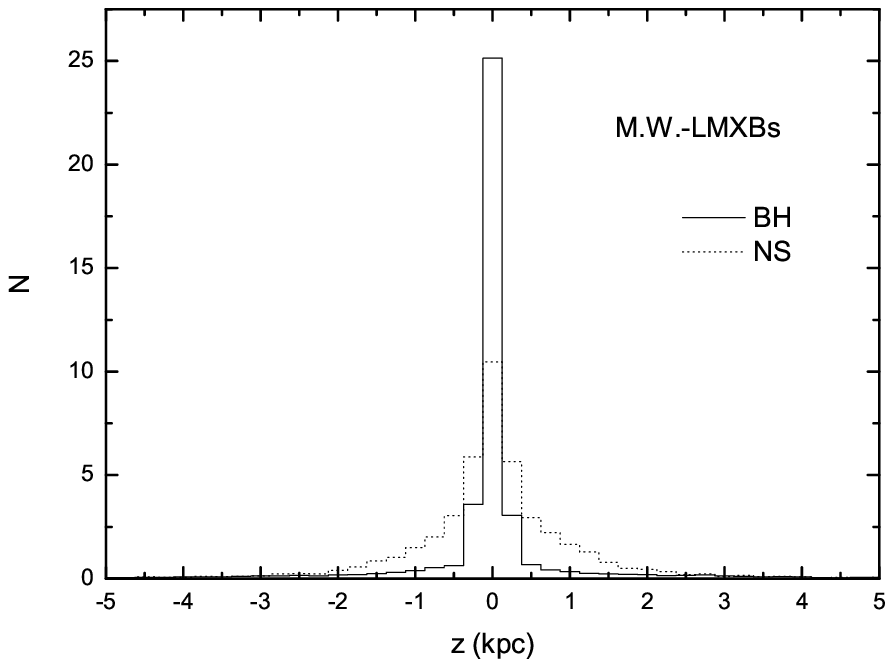}
  \centering
  \includegraphics[width=8cm,height=8cm]{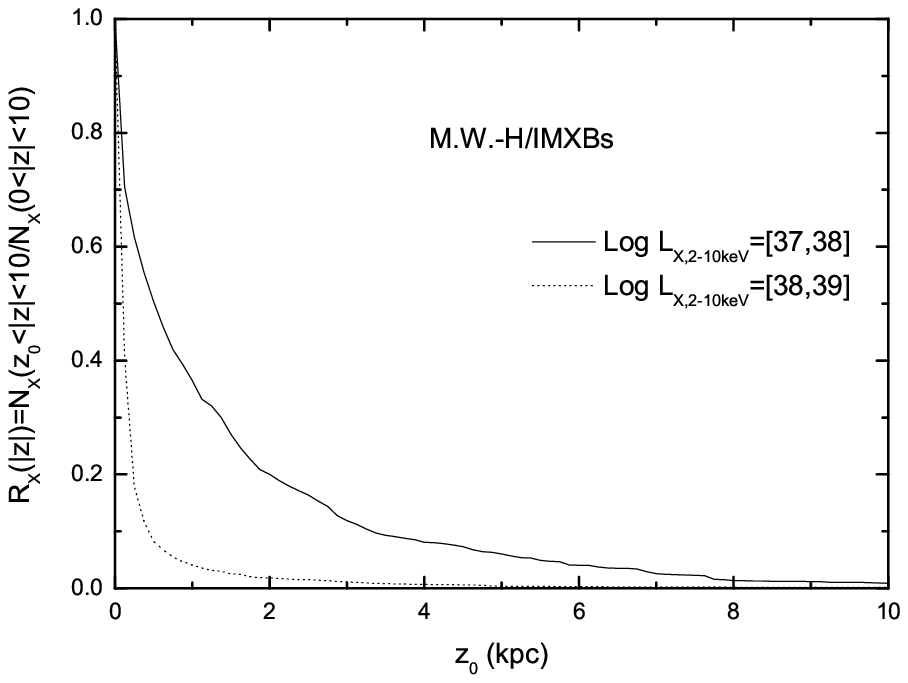}
  \caption{{\it Left panel:} Vertical distribution of LMXBs with BH (solid line) and NS (dotted line) accretors in the
Milky Way. {\it Right panel:} Variation of the ratio $R_{\rm
X}(|z|)$ for H/IMXBs in luminosity intervals $\log(L_{\rm
X,2-10kev})=[37,38]$ (solid line) and $[38,39]$ (dotted line) with
$z_0$ in the Milky Way.}
  \label{Fig. 4}
\end{figure}

\newpage
\begin{figure}
  \centering
  \includegraphics[width=8cm,height=8cm]{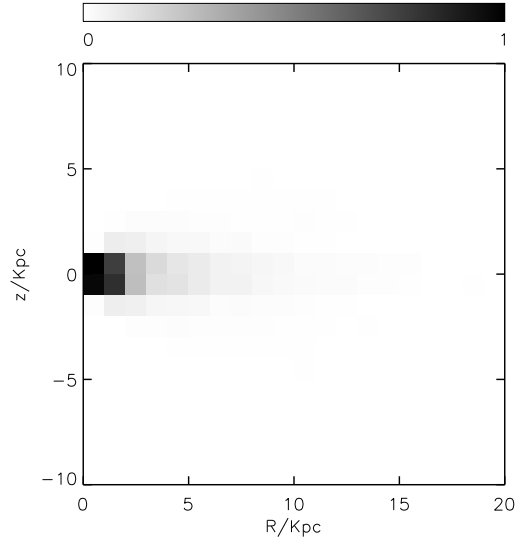}
  \caption{ The spatial distribution of luminous XRBs in the Milky Way in the CDM potential.
  The origin of the coordinate is at the Galactic center, $R$ and $z$ are
  the horizontal distance from the Galactic center and vertical distance from the Galactic plane in units of kpc.
  The colorbar represents the normalized number of XRBs.}
  \label{Fig. 5}
\end{figure}

\begin{figure}
  \centering
  \includegraphics[width=8cm,height=8cm]{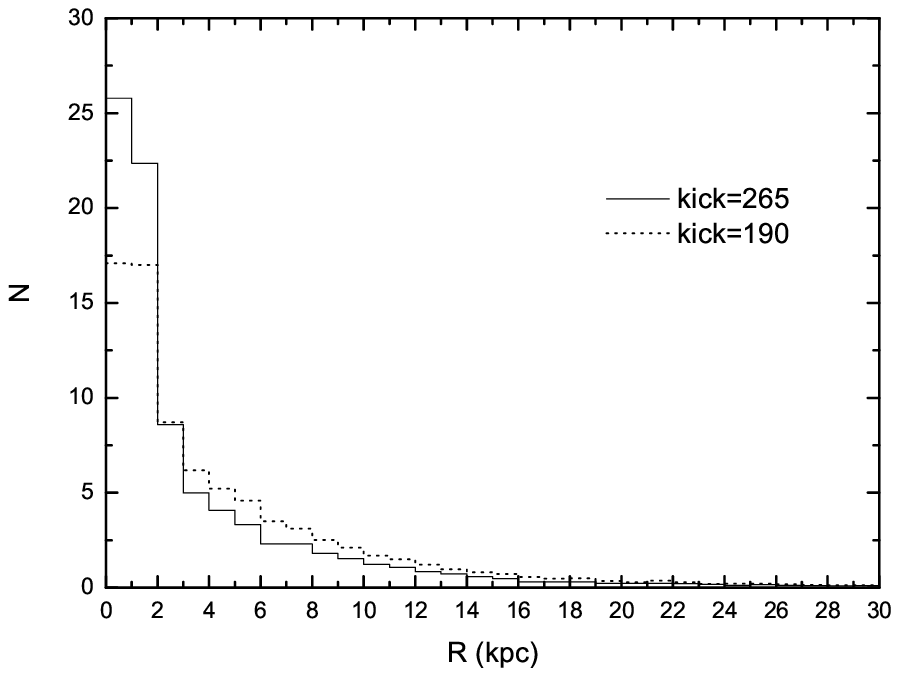}
\quad
  \includegraphics[width=8cm,height=8cm]{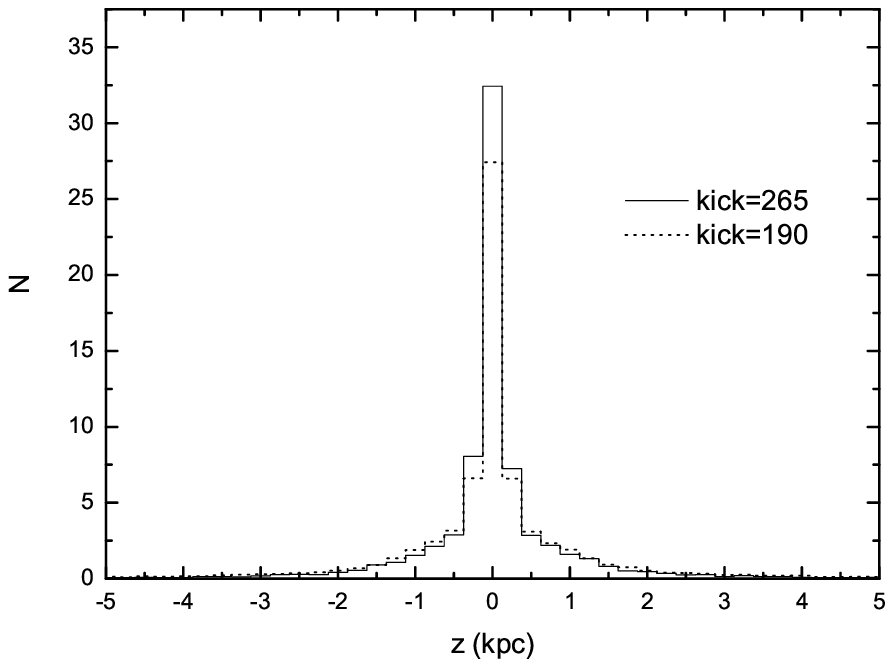}

  \caption{Dependence of the radial (left) and vertical (right) distribution on kick velocity dispersion of LMXBs in the
  Milky Way. The solid and dotted lines correspond to models with $\sigma=265{\rm kms}^{-1}$ \citep{Hobbs} and
  $\sigma=190{\rm kms}^{-1}$ \citep{hansen97}, respectively.}
  \label{Fig. 6}
\end{figure}

\begin{figure}
  \centering
  \includegraphics[width=8cm,height=8cm]{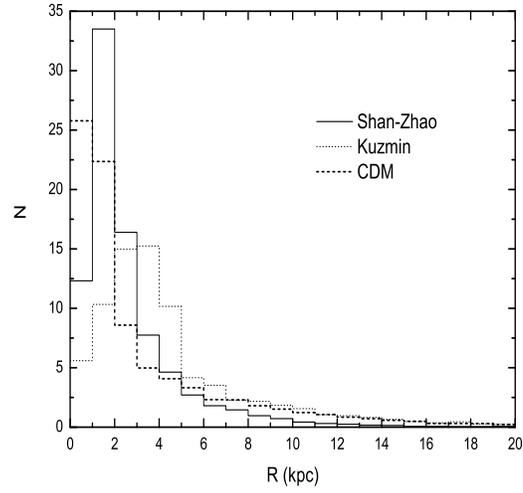}

  \caption{Radial distributions of XRBs with $L_{\rm x}>10^{37}$ergs$^{-1}$ in the Milky Way in MOND and CDM potentials.
  The solid, dotted, and thick short dashed lines represent the results in the Shan-Zhao model, Kuzmin model and CDM model, respectively.
 }
  \label{Fig. 7}
\end{figure}

\begin{figure}
  \centering
  \includegraphics[width=8cm,height=8cm]{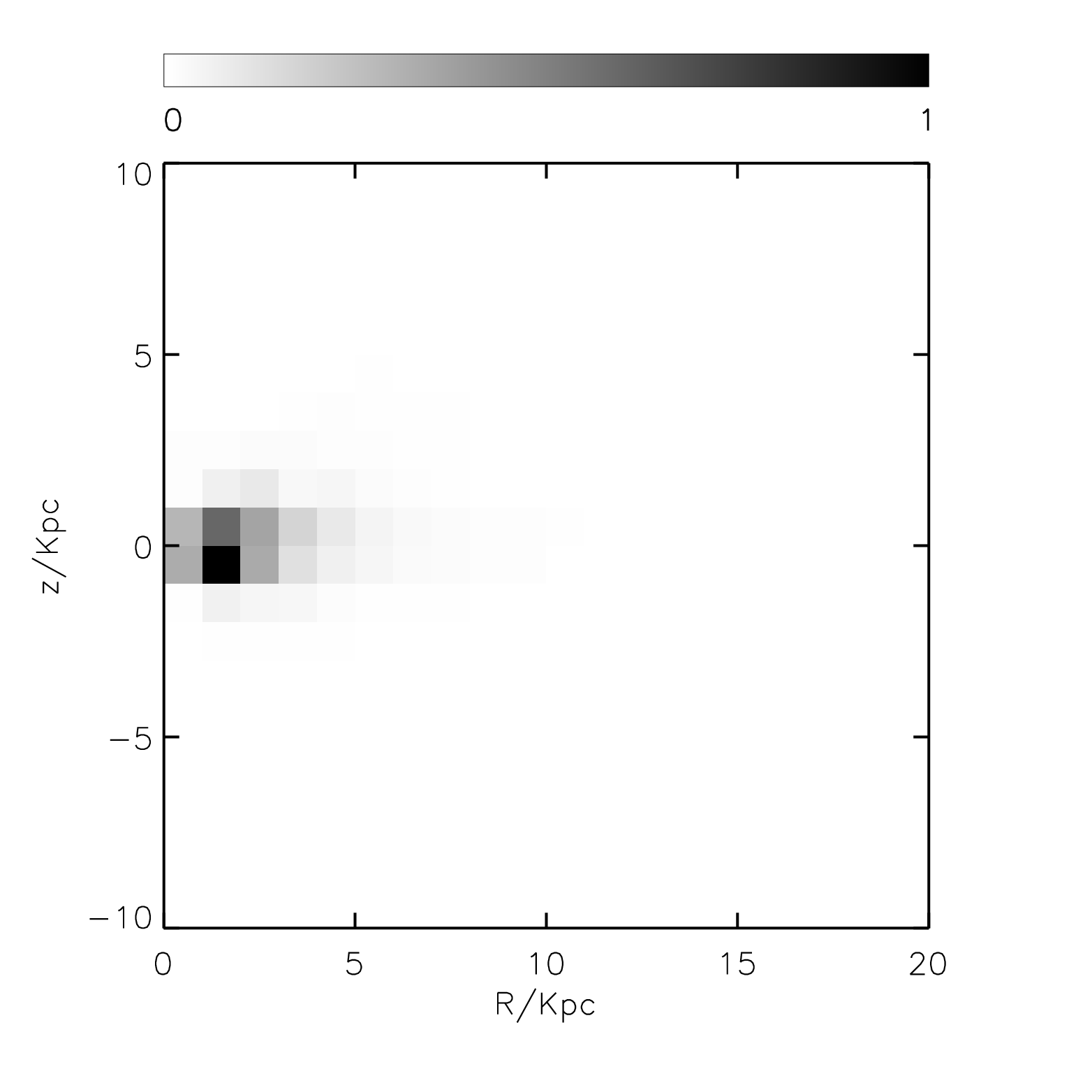}
\quad
  \includegraphics[width=8cm,height=8cm]{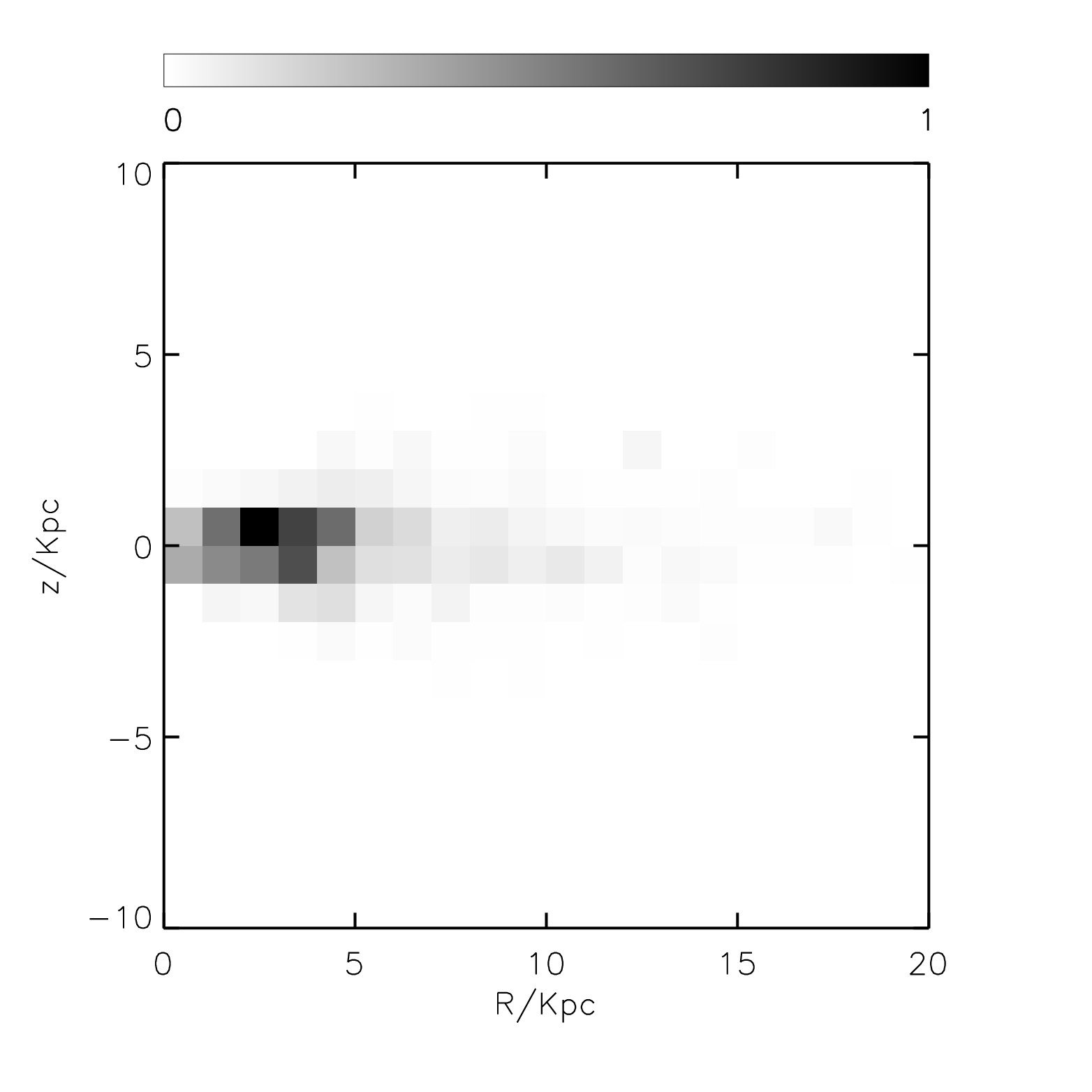}

  \caption{Same as in Figure 5, but for the MOND potential with Shan-Zhao model (left) and Kuzmin model (right).}
  \label{Fig. 8}
\end{figure}

\end{document}